\documentclass[preprint,10pt,5p,twocolumn,authoryear]{elsarticle}

\usepackage{amssymb}

\usepackage{hyperref}
\usepackage{relsize}

\journal{Astronomy and Computing}

\begin{document}

\def\apj{ApJ, }
\def\aj{AJ, }
\def\apjl{ApJL, }
\def\procspie{Proc.~SPIE}
\def\aap{A\&A, } 
\def\mnras{MNRAS, }
\def\araa{ARAA, }

\begin{frontmatter}



\title{fcmaker: automating the creation of ESO-compliant finding charts for Observing Blocks on \textit{p2}}


\author{Fr\'ed\'eric P.A. Vogt\corref{cor1}}
 \cortext[cor1]{ESO Fellow}
\ead{frederic.vogt@alumni.anu.edu.au}
\ead[url]{http://fpavogt.github.io}


\begin{abstract}
\textsc{fcmaker} is a \textsc{python} module that creates astronomical finding charts for Observing Blocks (OBs) on the \textit{p2} web server from the European Southern Observatory (ESO). It provides users with the ability to automate the creation of ESO-compliant finding charts for Service Mode and/or Visitor Mode OBs at the Very Large Telescope (VLT). The design of the \textsc{fcmaker} finding charts, based on an intimate knowledge of VLT observing procedures, is fine-tuned to best support night time operations. As an automated tool, \textsc{fcmaker} also provides observers with the means to independently check visually the observing sequence coded inside an OB. This includes, for example, the signs of telescope and position angle offsets. VLT instruments currently supported by \textsc{fcmaker} include MUSE (WFM-AO, WFM-NOAO, NFM), HAWK-I (AO, NOAO), and X-shooter (full support). The \textsc{fcmaker} code is published on a dedicated Github repository under the GNU General Public License, and is also available via \textsc{pypi}.

\end{abstract}


\end{frontmatter}


\section{Introduction}
Observing Blocks (OBs) are at the core of night time operations at the Very Large Telescope (VLT). They allow observers to define detailed observing sequences for all the instruments of the Unit Telescopes (UTs) on Cerro Paranal, as well as for the VLT Survey Telescope \citep[VST;][]{Arnaboldi1998}, the Visible and Infrared Survey Telescope for Astronomy \citep[VISTA;][]{Emerson2006}, and the VLT Interferometer (VLTI) instruments. In particular, OBs allow observers to specify the instrument observing mode, any optical element -- such as filters and grisms -- to be inserted in the optical path, the target coordinates, and the full observing sequence, including detector integration times, telescope offsets and position angle changes. To start an observation sequence, OBs are fetched by the night astronomer (NA) or telescope \& instrument operator (TIO) and sent to BOB\footnote{Broker for Observing Blocks}, a piece of software that parses the content of OBs and dispatches it to the appropriate systems: for example, the target coordinates are sent to the telescope, and the observing mode is sent to the instrument. Once the target acquisition is completed, BOB automatically feeds the details of each step within the observing sequence to the necessary systems. 

For Service Mode programs, observers are required to attach dedicated finding chart(s) to every one of their OBs. These finding charts are used by the NA and/or TIO at the VLT to confirm the correct acquisition of the target, and check that the execution of the OB proceeds as intended. For most instruments, finding charts are thus an essential component of night time operations.

Recently, the European Southern Observatory (ESO) released \textit{p2}\footnote{\url{http://www.eso.org/p2}, accessed on \today.}, a web application to support the preparation of OBs for Service Mode and Visitor Mode programs. \textit{p2} is designed to replace the standalone desktop tool \textit{p2pp}. As of the observing period P101, observers with Visitor Mode programs on any UT, and observers with Service Mode programs on X-shooter \citep{Vernet2011}, FLAMES \citep[Fibre Large Array Multi Element Spectrograph;][]{Pasquini2002}, UVES \citep[Ultraviolet and Visual Echelle Spectrograph;][]{Dekker2000}, OmegaCAM \citep[][]{Kuijken2002,Kuijken2011} and VIRCAM \citep[VISTA InfraRed CAMera;][]{Dalton2006} (i.e. all UT2 instruments and the survey telescopes) are asked to use \textit{p2} for preparing their OBs. The transition from \textit{p2pp} to \textit{p2} for Service Mode programs on all other VLT instruments occurred ahead of P102.

Alongside \textit{p2}, ESO also released \textsc{p2api}: a dedicated programming interface (API) in the form of a \textsc{python} module \citep{Bierwirth2018}. This module opens the door to automating a range of operations on \textit{p2}: for example, the mass production of generic OBs, a feat largely reserved (in the pre-\textit{p2} era) to public surveys on VST and VISTA. This article presents \textsc{fcmaker}\footnote{DOI:10.5281/zenodo.1163010}\footnote{\url{https://ascl.net/1806.027}}, a \textsc{python 3} module that exploits the capabilities of \textsc{p2api} to automatically create ESO-compliant finding charts for existing OBs on \textsc{p2}. The design drivers of \textsc{fcmaker} are discussed in Sec.~\ref{sec:design}, and the actual implementation of the tool (including its output) is described in Sec.~\ref{sec:fcmaker}. The planned mid-term future for \textsc{fcmaker} is presented in Sec.~\ref{sec:plans}. It is important to note that \textsc{fcmaker} is not an official ESO tool. It is free software, published under the terms of the GNU General Public License as published by the Free Software Foundation, version 3 of the License. As such, \textsc{fcmaker} is distributed in the hope that it will be useful, but without any warranty; without even the implied warranty of merchantability or fitness for a particular purpose.

\section{The design drivers of fcmaker}\label{sec:design}

An OB, when complete, contains all the necessary information required to create the associated finding chart. This was already true in the pre-\textit{p2} era, but it is the new capabilities of this web interface, and in particular its dedicated API, that today allows this fact to be exploited with \textsc{fcmaker}. When compared to a manual approach, the automated creation of finding charts for OBs on \textit{p2} presents several advantages. First, it can allow for the batch processing of numerous OBs efficiently, without any further (redundant) input from the observer. Second, it offers observers the means to check the content of their OBs: for example, the signs of blind offsets and position angle changes. Without automated finding chart(s), wrong signs or typos inside OBs may remain undetected up until the execution of the OB at the VLT, and the realization by the operator that the OB acquisition/execution does not match the submitted finding charts. Third, automated finding charts can be made to always fulfill all of ESO's formal requirements\footnote{\url{https://www.eso.org/sci/observing/phase2/SMGuidelines/FindingCharts.html}, accessed on \today.}. These generic finding chart requirements, applicable to all of the VLT instruments, are as follows:
\begin{itemize}
\setlength\itemsep{0pt}
 \item clearly indicate the Observing Run identification (ID) number;
 \item clearly indicate the primary investigator (PI) name;
  \item clearly indicate the OB name or Target name, as used in the OB;
 \item clearly indicate the target(s) position(s);
 \item the entire instrument field-of-view must be shown;
 \item clearly indicate North and East;
 \item the scale must be indicated by drawing a bar and writing the bar length in arcseconds or arcminutes;
 \item the wavelength range of the image must be indicated. Whenever possible, finding charts should have similar central wavelength to observations (e.g. DSS charts are often inappropriate for IR observations near the galactic equator);
 \item the images should be negative, i.e. dark objects on light background;
 \item the output files must be in JPEG format and their size must be less than 1 Mbyte;
 \item positions of spectroscopic acquisition reference stars, if any, should be marked;
 \item spectroscopic finding charts must indicate the slit(s) position(s) clearly (unless slits are aligned along the parallactic angle).
\end{itemize}

\textsc{fcmaker} is build to meet all of these requirements, and more. It aims at providing clear finding charts fine-tuned for supporting night time operations at the VLT. This philosophy implies that at times, finding charts generated by \textsc{fcmaker} may not be optimized for \textit{preparing} OBs. For example, \textsc{fcmaker} finding charts for MUSE WFM OBs do not show the Slow-Guiding-System (SGS) area (see Fig.~\ref{fig:MUSE}). Ensuring that a star falls within the MUSE SGS area for all offsets in a given observing sequence may be an important OB design goal, but the actual selection of stars within the SGS area during the execution of an OB is fully automated: drawing the SGS area on finding charts is thus unnecessary for operations. From that perspective, \textsc{fcmaker} is complementary to OB preparation tools such as ESO's GUideCam Tool (GUCT)\footnote{\url{https://www.eso.org/sci/observing/phase2/SMGuidelines/GUCT.generic.html}, accessed on \today.}.

\textsc{fcmaker} was created by an ESO Fellow with duties on UT4. Beyond ESO's formal requirements, the design of the \textsc{fcmaker} finding charts is thus also driven by an intimate knowledge of night time operations at the VLT. As a consequence, certain design choices for the \textsc{fcmaker} finding charts voluntarily exceed ESO's formal requirements, in order to better support operations. For example, \textsc{fcmaker} finding charts clearly flag telescope guide stars compatible with all telescope offsets within a given observing sequence. As \textsc{fcmaker} is fully automated, these enhanced features are directly available ``for free'' to the user. 

\textsc{fcmaker} is (voluntarily) providing users with very limited options, including the choice of the background image, and the time of the observation. By default, \textsc{fcmaker} generates a unique finding chart per OB. The use of \textsc{fcmaker} thus does not preclude observers from attaching additional finding charts to their OBs, should their observing requirements warrant it\footnote{OBs can currently accommodate a maximum of 5 finding charts.}. In particular, the use of GUCT does not prevent the use of \textsc{fcmaker}, and vice-versa.

\section{The implementation of fcmaker}\label{sec:fcmaker}

\textsc{fcmaker} is written in \textsc{python 3}. The code is hosted on a dedicated Github repository\footnote{\url{https://github.com/fpavogt/fcmaker/}}, of which the releases\footnote{\url{https://github.com/fpavogt/fcmaker/releases}} are all assigned their own Digital Object Identifier (DOI) via Zenodo\footnote{All versions of \textsc{fcmaker} can be cited using \href{https://doi.org/10.5281/zenodo.1163010}{DOI:10.5281/zenodo.1163010}, which will always resolve to the latest one.}. Every stable release of \textsc{fcmaker} is available on \textsc{pypi}, which, in turn, provides users with the means to install the module with a typical: \texttt{pip install fcmaker}. Detailed installation and usage instructions are included in the online documentation\footnote{\url{hhtp://fpavogt.github.io/fcmaker}}. These are voluntarily not repeated here in detail, since they might be evolving over time. It suffices to say that in its current implementation, \textsc{fcmaker} 1) uses the \textsc{p2api} package to retrieve the content of OBs from \textit{p2}, 2) creates the associated finding charts locally, and 3) uses the \textsc{p2api} package to attach the resulting image files to the OBs on \textit{p2}. Higher-level functions within \textsc{fcmaker} can be imported inside custom \textsc{python} scripts. Alternatively, the module itself is also designed to be run as a script using the command: \texttt{python -m fcmaker}. The latter approach ought to be more user-friendly to users lacking experience with \textsc{python}. In addition to the creation of finding charts for OBs on \textit{p2}, \textsc{fcmaker} also contains a \textit{local} mode: this allows the creation of finding charts from custom parameter files stored locally. Users are encouraged to report any unexpected behaviour of \textsc{fcmaker} on the ``Issues'' page of the Github repository\footnote{\url{https://github.com/fpavogt/fcmaker/issues}}.

\begin{figure*}[htb!]
\vspace{40pt}
\centerline{\includegraphics[width=\textwidth]{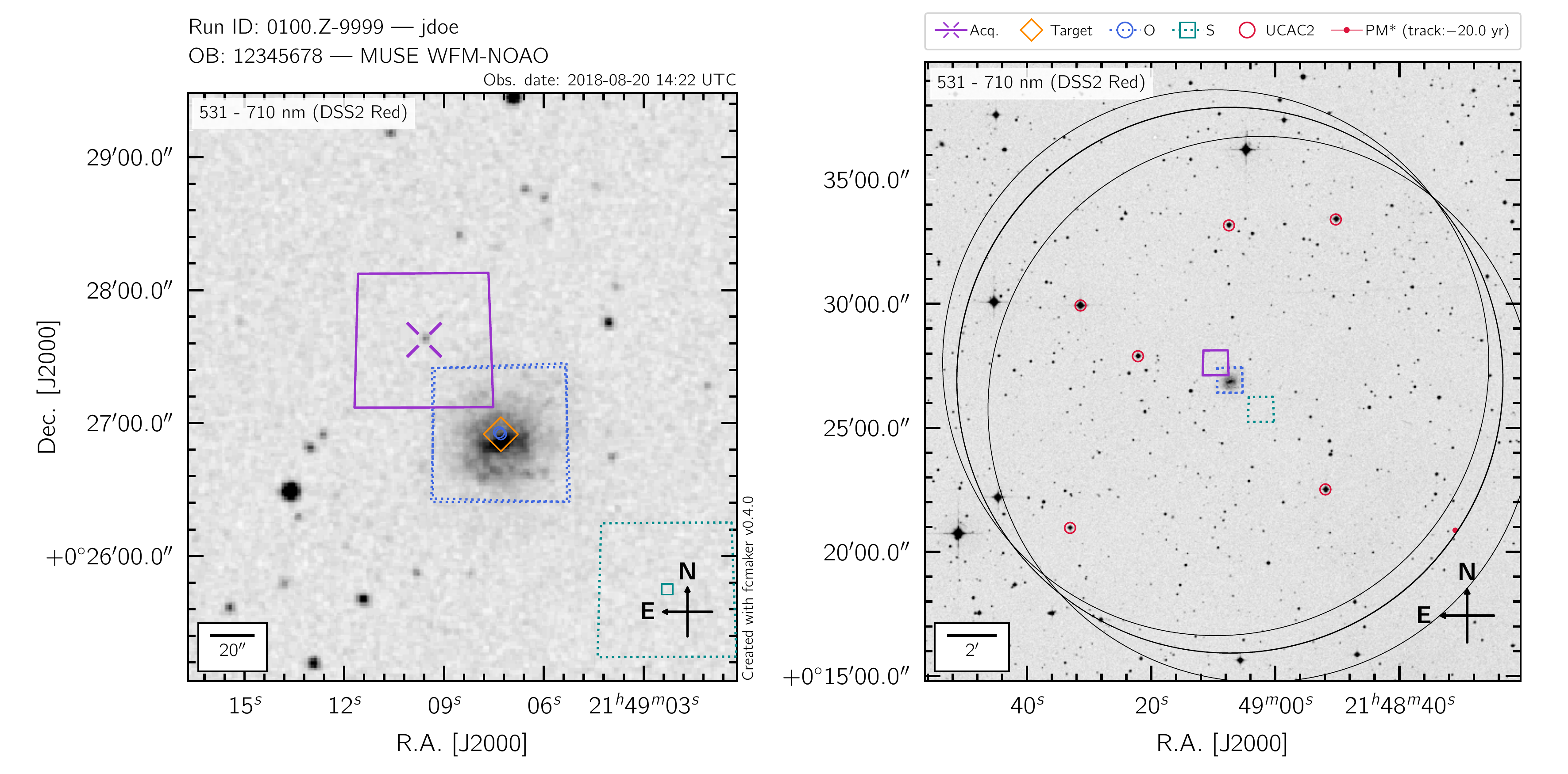}}
\vspace{30pt}
\centerline{\includegraphics[width=\textwidth]{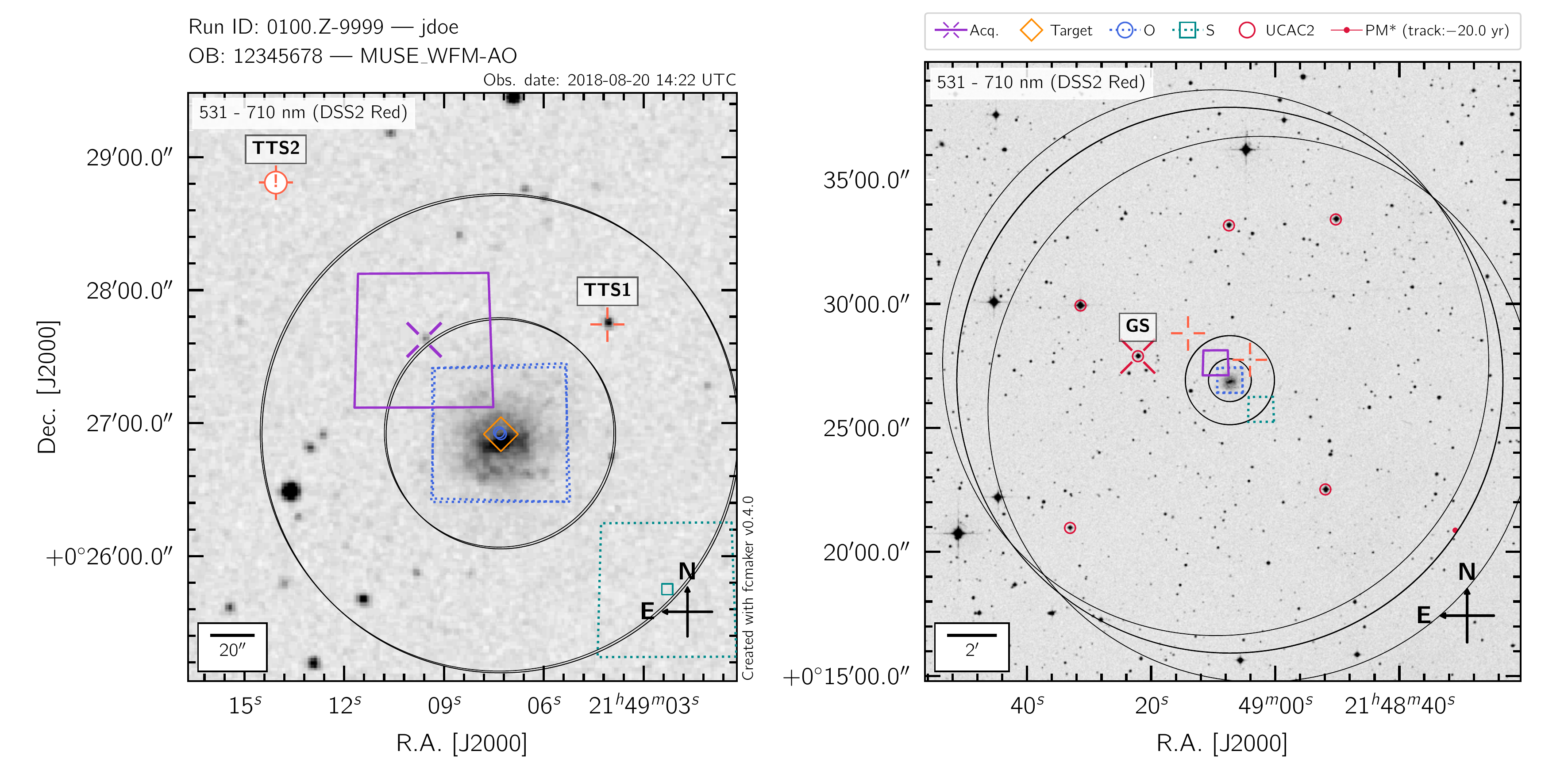}}
\caption{Examples of \textsc{fcmaker} finding charts for MUSE WFM-NOAO (top) and WFM-AO (bottom) OBs. The module creates a single image composed of two sup-panels per OB: the right-hand-side panel panel provides a large view of the telescope guide star selection area, whereas the left-hand-side panel present a smaller view focusing on the acquisition field (in bold purple). Suitable telescope guide stars (meeting the nominal selection criteria) are clearly marked with red circles on the right-hand-side, and so are the different object (O) and sky (S) fields within the OB (dotted lines). For MUSE WFM-AO, the validity area for the Tip-Tilt Stars (TTS) is indicated, alongside with any TTS specified by the user in the OB. User-defined telescope guide stars (GS) are drawn accordingly. }\label{fig:MUSE}

\end{figure*}

\begin{figure*}[htb!]
\vspace{-20pt}
\centerline{\includegraphics[width=0.8\textwidth]{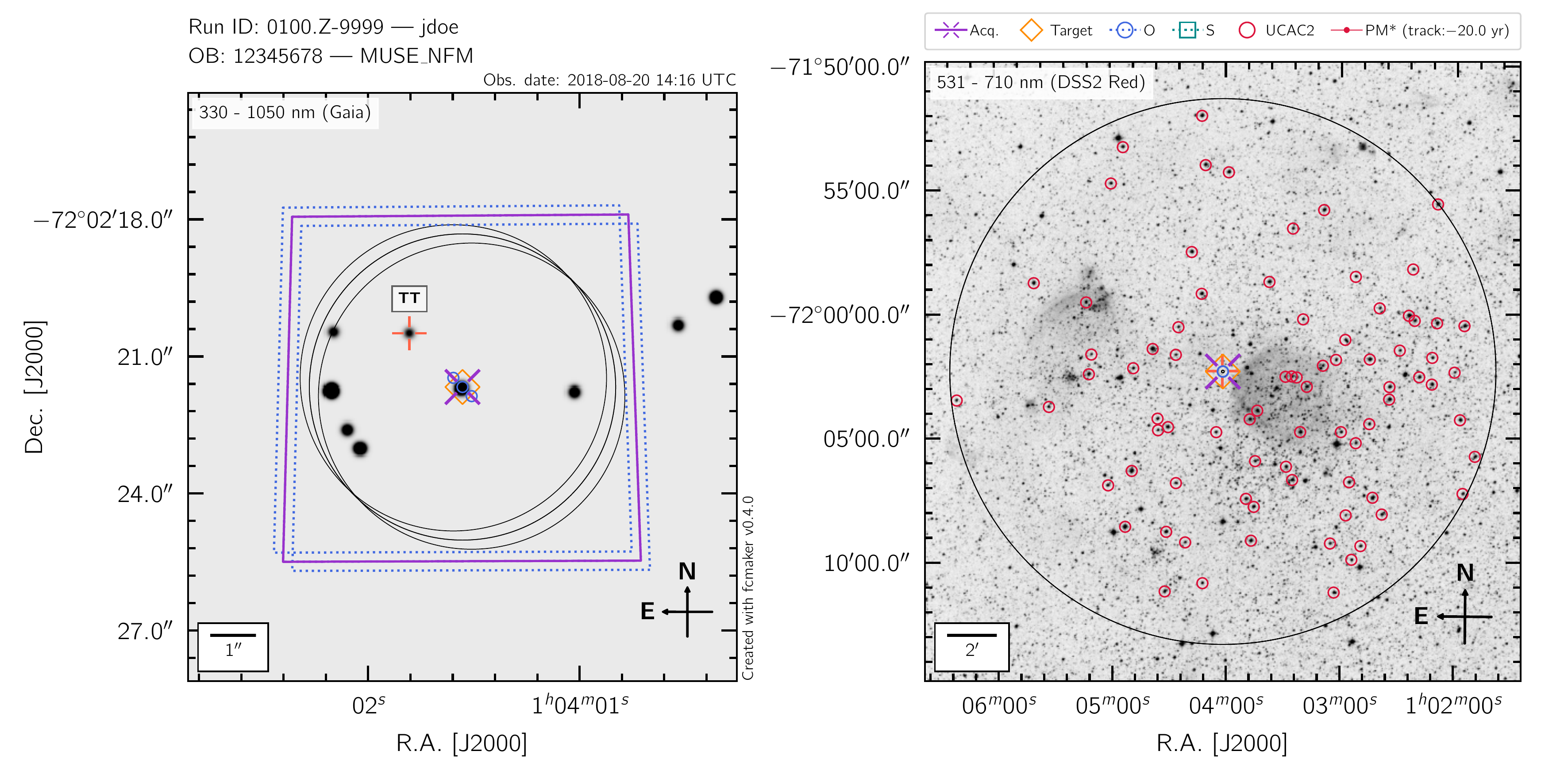}}
\vspace{5pt}
\centerline{\includegraphics[width=0.8\textwidth]{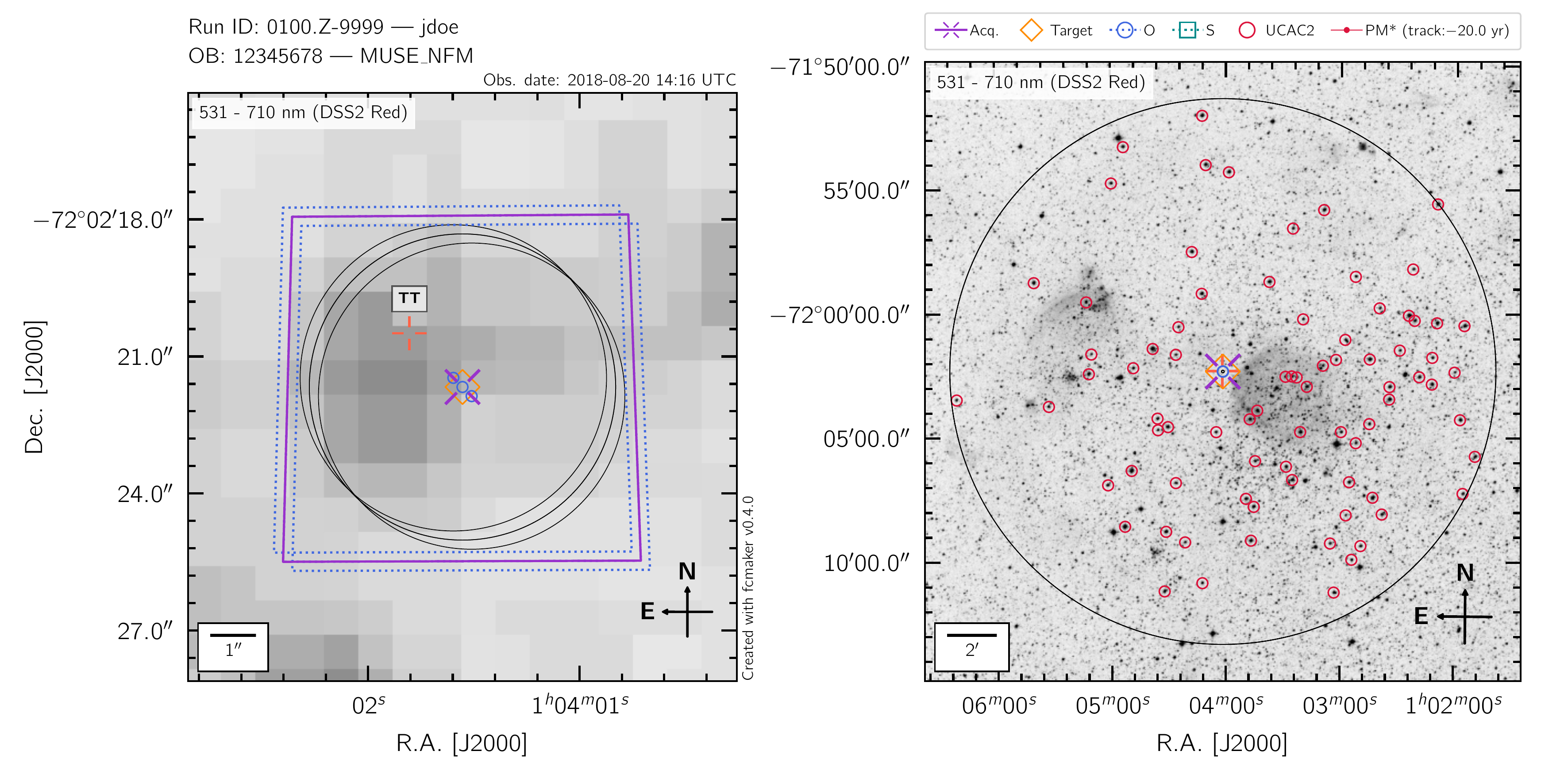}}
\vspace{5pt}
\centerline{\includegraphics[width=0.8\textwidth]{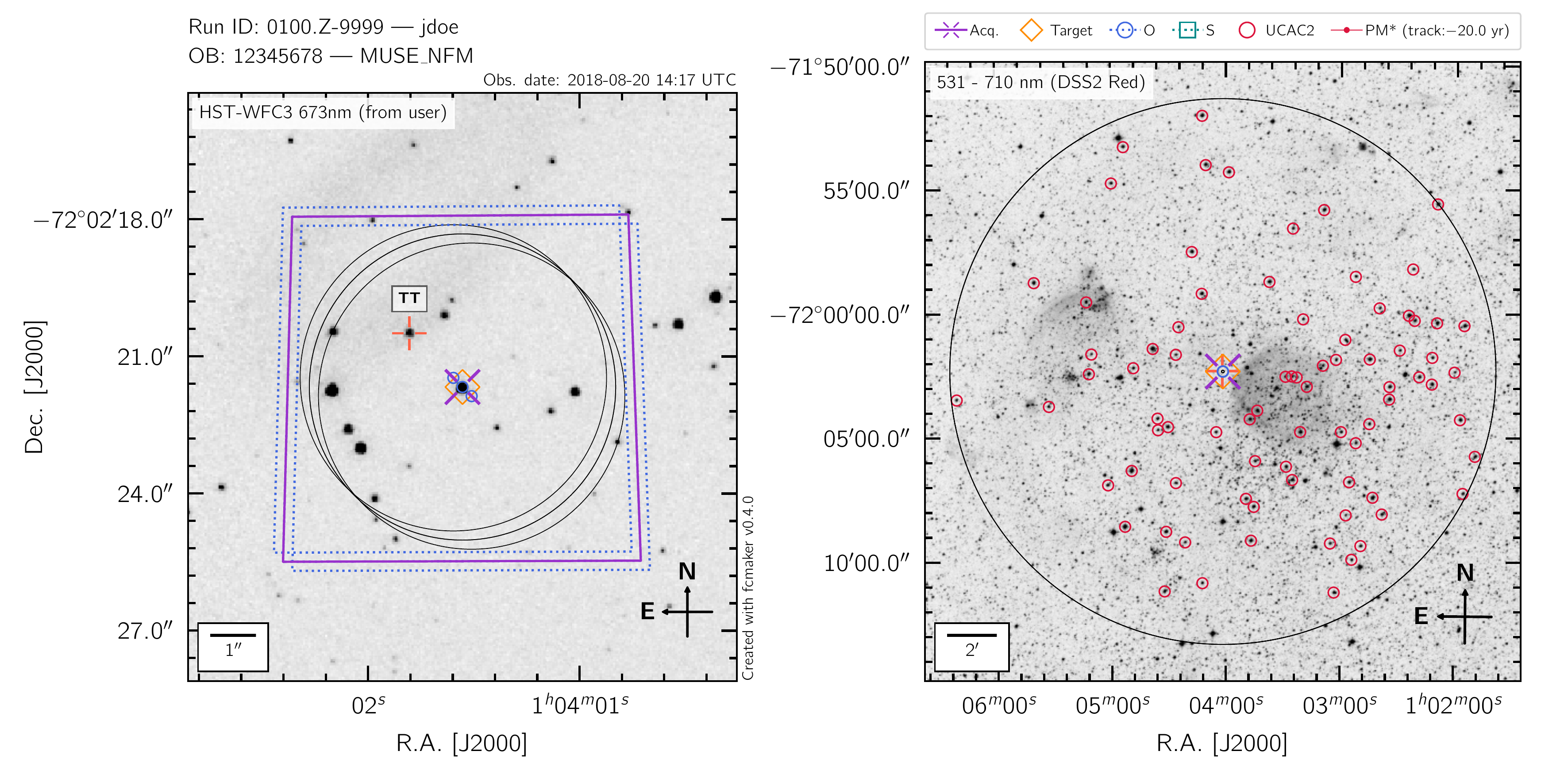}}
\caption{Example of \textsc{fcmaker} finding charts for MUSE NFM. The top plot demonstrates the default behavior of fcmaker, which constructs (for the left-hand-side panel) a mock sky image based on the entries in the \textit{Gaia} catalogue over the area. Stars are shown as 2D gaussians, scaled in intensity as a function of their \textit{Gaia} magnitude, and with a full-width-at-half-maximum of 80 milli-arcsec (by default). In the case of MUSE NFM, such mock sky images are significantly more useful than, for example, the DSS2 red image of the same area (middle). Evidently, the best case scenario is when \textit{real} high resolution images of the area are available (bottom), for example from the Hubble Space Telescope (HST). Any FITS image with proper WCS information can be fed by the user to \textsc{fcmaker} for the left-hand-side plot background.}\label{fig:MUSE-NFM}
\vspace{0pt}
\end{figure*}

\begin{figure*}[htb!]
\vspace{40pt}
\centerline{\includegraphics[width=\textwidth]{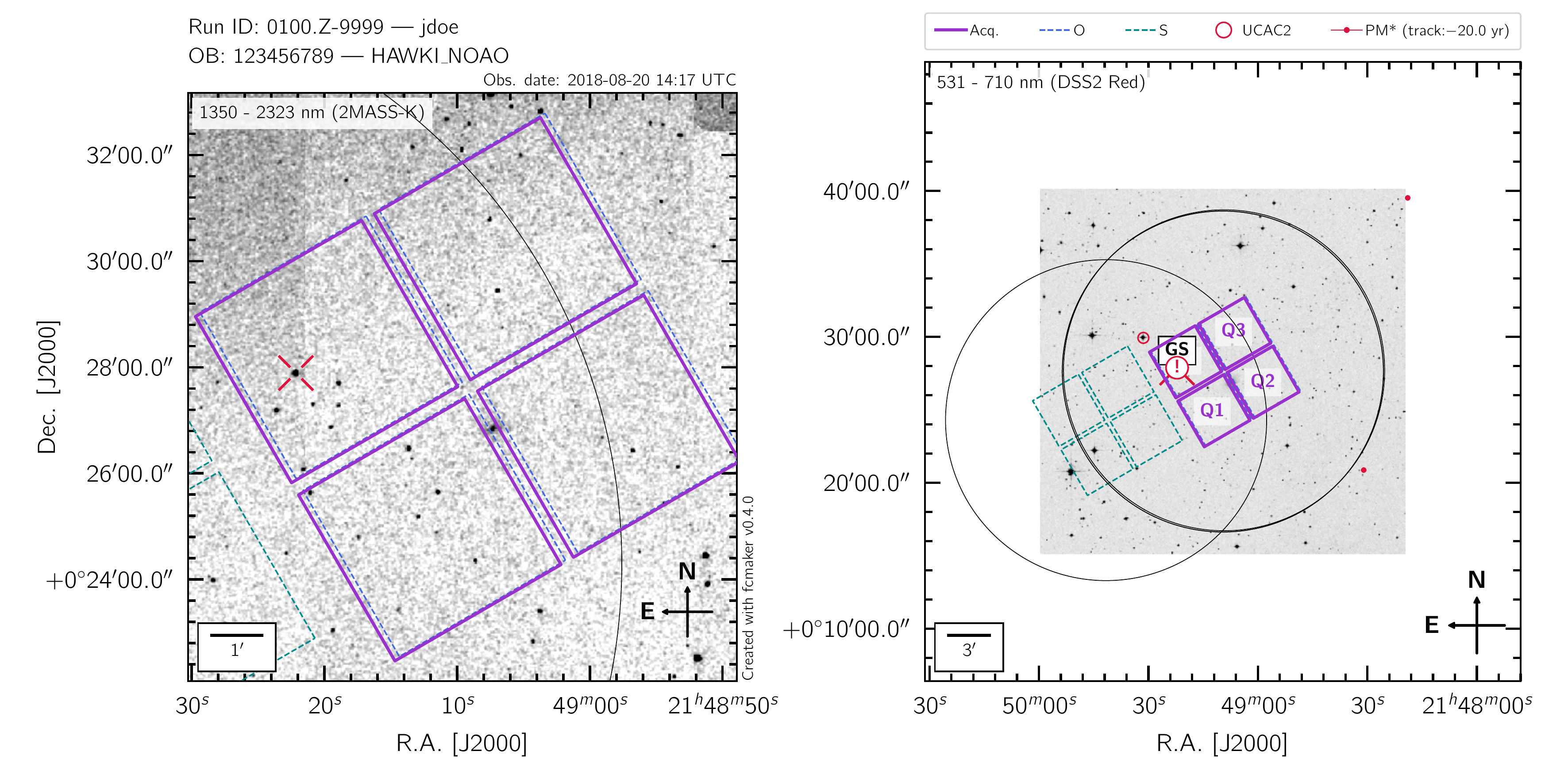}}
\vspace{30pt}
\centerline{\includegraphics[width=\textwidth]{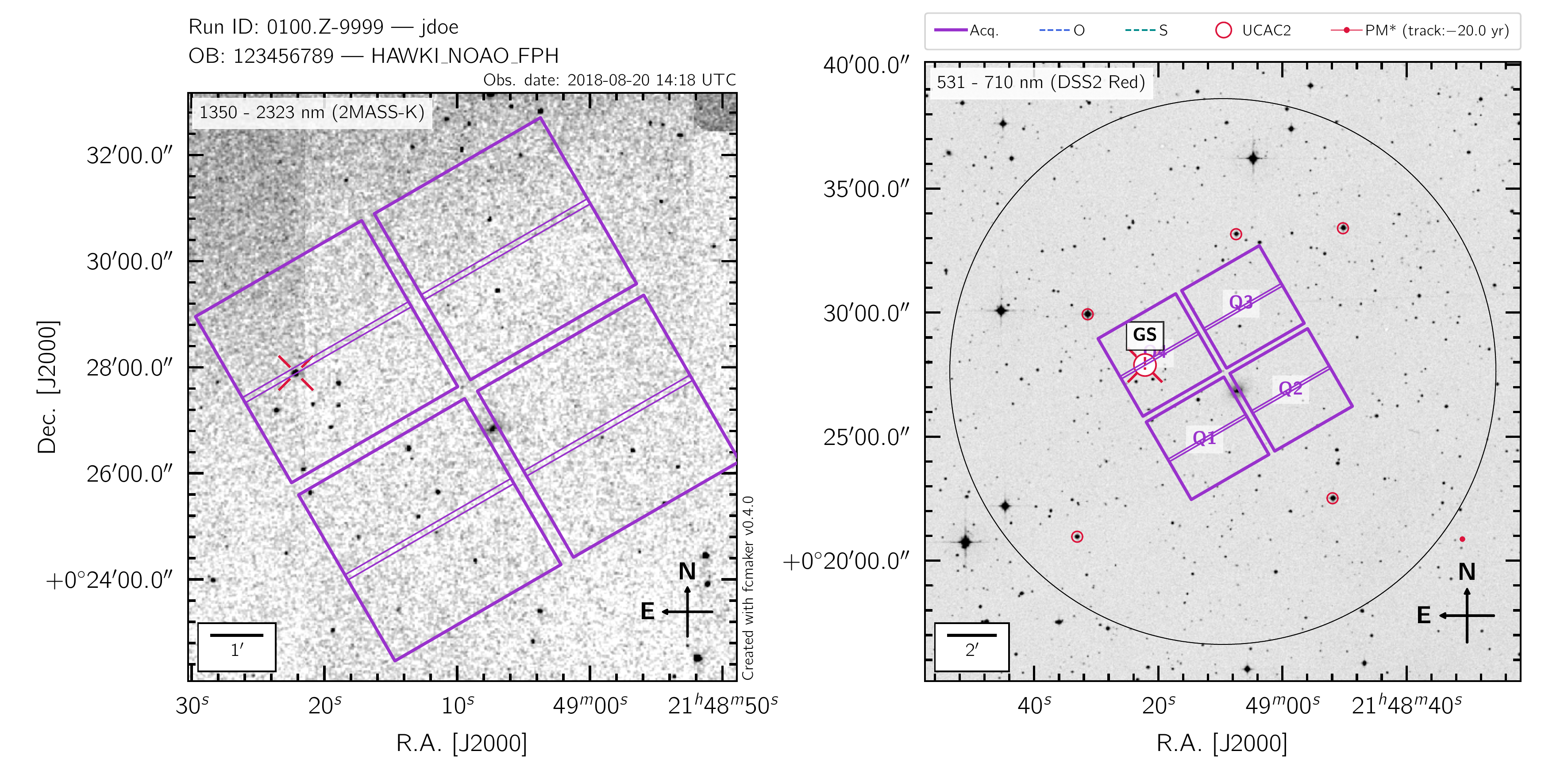}}
\caption{Example of \textsc{fcmaker} finding charts for HAWK-I NOAO (top) and HAWK-I NOAO fast-photometry (bottom) OBs. Since HAWK-I AO is currently offered only in tip-tilt-free mode, the finding charts associated with this mode are identical to the NOAO ones. In case of fast-photometry observations, the detector read-out areas specified in the OB are drawn accordingly.}\label{fig:HAWKI}
\vspace{50pt}
\end{figure*}

\begin{figure*}[htb!]
\vspace{40pt}
\centerline{\includegraphics[width=\textwidth]{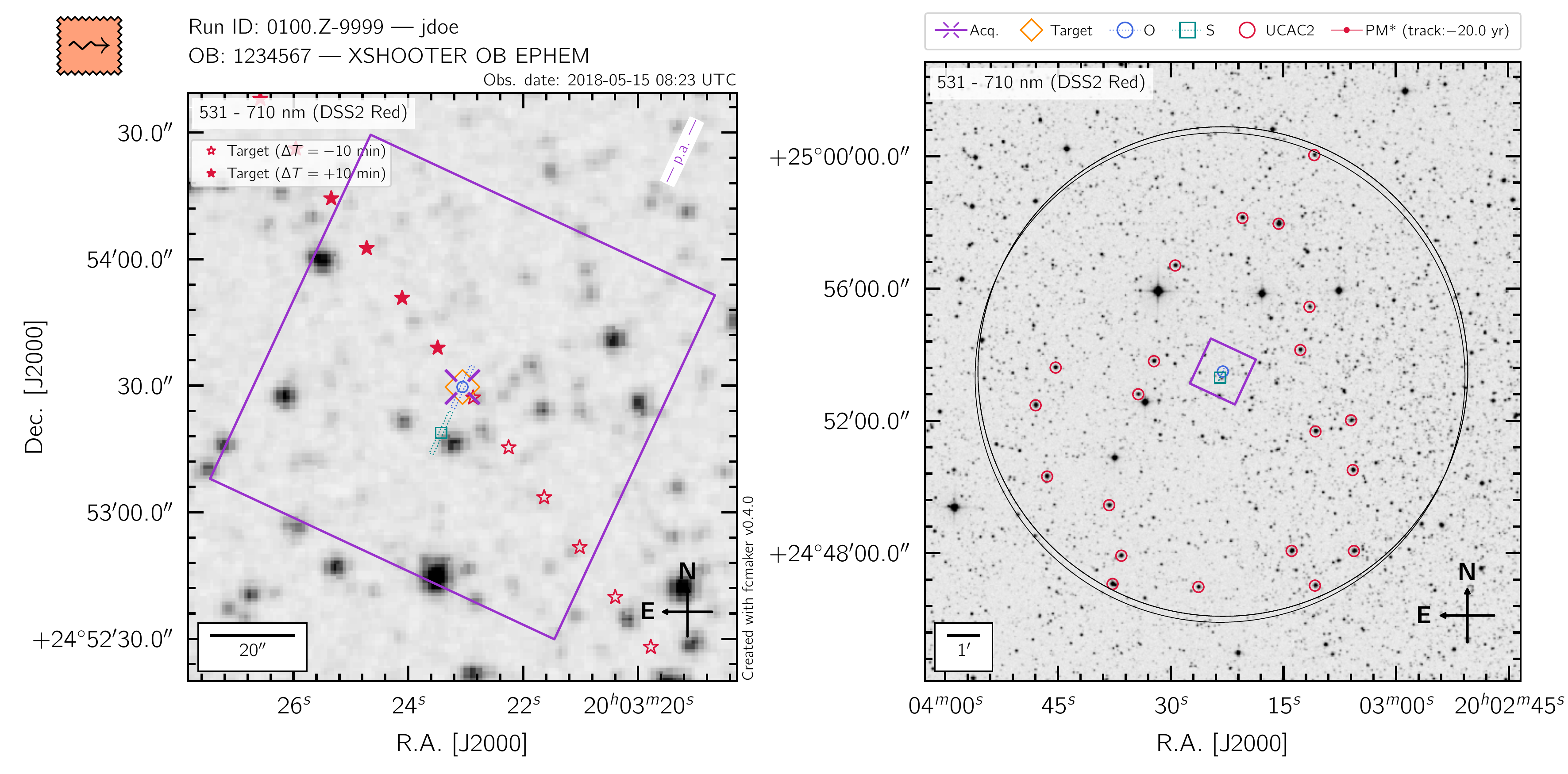}}
\vspace{30pt}
\centerline{\includegraphics[width=\textwidth]{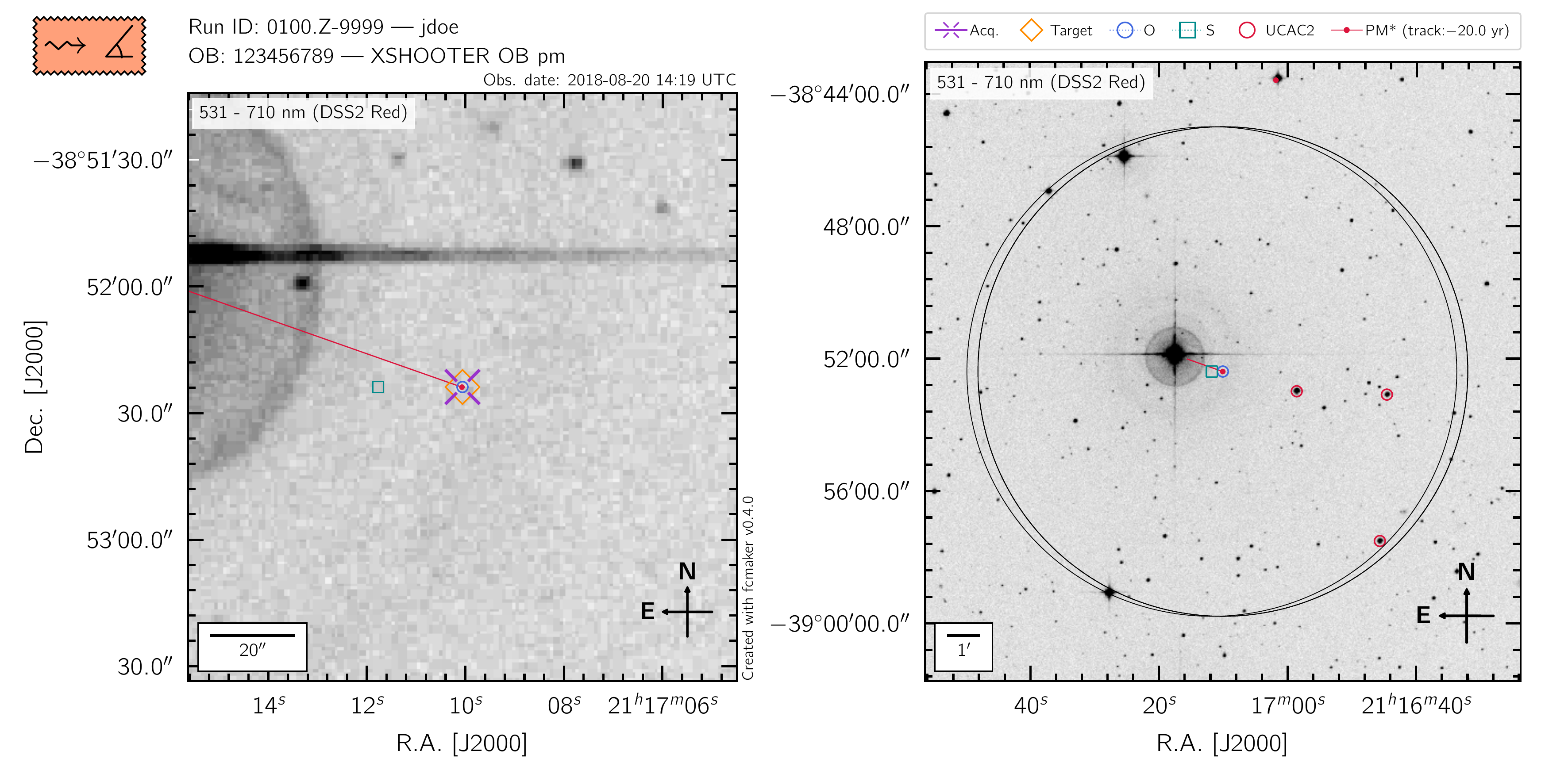}}
\caption{Example of \textsc{fcmaker} finding charts for X-shooter. The field-of-view of the acquisition camera is shown in bold purple -- except when a parallactic angle is being requested. The top chart illustrates how moving targets, defined using an ephemeris file, are treated. Each ephemeris point with $\pm$2\,h (by default) from the observing time specified by the user are drawn individually using red stars. The bottom chart illustrates how the observation of a moving target requesting a parallactic angle is treated: only the field centers for the acquisition, object (O) and sky (S) positions are drawn, but not the full field-of-views, to avoid rendering the finding chart highly time sensitive.}\label{fig:XSHOOTER}
\end{figure*}

Example finding charts for the (currently) supported instruments -- MUSE \citep[Multi-Unit Spectroscopic Explorer;][]{Bacon2010}, HAWK-I \citep[High-Acuity Wide-field K-band Imager;][]{Pirard2004,Casali2006,Kissler-Patig2008,Siebenmorgen2011}, and X-shooter -- are presented in Fig.~\ref{fig:MUSE} to \ref{fig:XSHOOTER}. They illustrate the design choices built into \textsc{fcmaker}.

All \textsc{fcmaker} finding charts share common traits. The run ID, PI name, OB number (if applicable) and OB name are all clearly indicated: so are the N and E directions, as well as the chart angular scale. The finding charts are single images containing two display panels. The right-hand-side panel is primarily aimed at the TIO. It shows a wide-field view of the observation area, together with the nominal search area for telescope guide stars (11$^{\prime\prime}$ in radius for Nasmyth instruments, 7.4$^{\prime\prime}$ in radius for Cassegrain instruments) for every pointing inside the observing sequence of the OB. All guide stars compatible with the nominal VLT requirements are marked with red circles. These are entries in the UCAC2 catalogue\footnote{UCAC2 is the catalogue currently used at the VLT for selecting telescope guide stars.} \citep{Zacharias2004} with a UCAC magnitude $11\leq UCACmag \leq 14$, and accessible by the guide probe for all the telescope offsets during the OB. It must be stressed that \textsc{fcmaker} does not check for the possible vignetting of the observations by the arm of the guide probe.

The background image for the right-hand-side plot is always taken from the Second Digitized Sky Survey (DSS2) red database, which matches the guiding wavelength of the UTs. Any star in the chart area with large proper motions (pm$\geq$100mas/yr as a default) measured by \textit{Gaia} \citep{GaiaCollaboration2016a} is flagged with a red dot, with its motion over the past 20 yr marked with a thin red line (see Fig.~\ref{fig:XSHOOTER} for an example). If specified, the user-selected guide star is marked with a red crosshair: if it falls outside of the valid telescope guide star zone for \textit{any} of the offset positions specified in the OB, it is flagged with the symbol ``\texttt{!}''. This is one of a very few courtesy checks performed by \textsc{fcmaker}, which will never replace the official OB validation mechanism built inside \textit{p2}. 

The left-hand-side plot is aimed at both the NA and TIO. It shows a detailed view of the observing region. The background image is instrument dependant, and can also be provided by the user. The acquisition field, which plays a crucial role to confirm/perform the successful acquisition of the OB, is always shown in bold purple. The center of this field is marked with a crosshair symbol, to highlight the location of the target (or blind offset star, if applicable). The other elements of the \textsc{fcmaker} finding charts are instruments dependant: they are discussed in the following subsections.

Unless provided by the user, the background images are downloaded automatically from the World Wide Web with the \textsc{astroquery} package \citep{Ginsburg2017}, which provides a single \textsc{python} entry point for a wide range of all-sky survey data, including DSS2, 2MASS, UCAC2, and \textit{Gaia}. The background images typically weigh 16\,MB, and require 2-3 minutes to be downloaded. The generation of the finding chart then requires an additional $\sim$20\,s, which is dominated by online queries to the UCAC2 and Gaia catalogues. Upon their download, the background images are stored in a local cache, so that re-generating a finding chart only requires $\sim$10\,s on a laptop. The weight of the resulting image files ranges from 300\,KB (when saved to the \textit{jpeg} format) to 600\,KB (when saved to the \textit{pdf} or \textit{png} formats).

\paragraph{MUSE}

The finding charts in WFM-NOAO mode (see Fig.~\ref{fig:MUSE} - top) show the location of the target defined in the OB, the acquisition field, and the subsequent object (O) and sky (S) fields. The crosshair marking the center of the acquisition field is not drawn for OBs that use the \texttt{MUSE\_wfm-noao\_acq\_preset} acquisition template, which does not require the operator to perform a fine-centering on the target (or blind offset star).

The minimum inner validity radius for telescope guide stars is 120$^{\prime\prime}$ from any offset position. The left-hand-side plot is always centered on the mid-point between the acquisition position and the target coordinates, with a radius sufficient to see both in the chart window. Depending on the telescope offsets in the OB, not all of the O or S fields may be visible in the left-hand-side plot. This design choice ensures that the acquisition field is always visible with a sufficient level of details. All the O and S fields are however always visible in the right-hand-side plot, to allow the NA and /or TIO to assess the correct execution of the OB. By default, the background image is taken from DSS2 (red).

In addition to these elements, the MUSE WFM-AO charts (see Fig.~\ref{fig:MUSE} - bottom) display the validity area for the tip-tilt stars for all OB offsets. This includes the target itself, which is when the AO loop is first closed, even if the first offset in the OB is not null. This however does not include the acquisition position, which implies that tip-tilt stars can be used as blind offsets in MUSE WFM-AO mode.  The first and second tip-tilt stars selected by the user are marked with red crosshairs. If any of the tip-tilt star falls outside of the validity area for any of the offset position, it is flagged with the symbol ``\texttt{!}''. This is another one of the very few courtesy checks performed by \textsc{fcmaker}. 

The MUSE NFM finding charts (see Fig.\ref{fig:MUSE-NFM}) display the validity area for the on-axis tip-tilt star for all OB offsets. If the on-axis tip-tilt star falls outside of the suitable area for any of the offset position, it is flagged with ``\texttt{!}''. By default, the background image for the left-hand-side panel of the MUSE NFM finding charts is a mock image of the sky, reconstructed from the \textit{Gaia} catalogue. This approach allows to assemble a useful view of the observation area despite the very small field-of-view of MUSE in this mode. \textsc{fcmaker} queries the latest \textit{Gaia} catalogue to construct this image, in which stars are shown as 2D gaussians, scaled in intensity as a function of their \textit{Gaia} magnitude, and with a full-width-at-half-maximum of 80 milli-arcsec (by default). The resulting image is saved as a FITS file with a fully-fledged header (including the World Coordinate System parameters) made available to the user. It typically weighs 2.2\,MB.

\paragraph{HAWKI}

The finding charts in NOAO mode (see Fig.~\ref{fig:HAWKI}) show the acquisition field, and the subsequent O and S fields if the \texttt{HAWKI\_img\_obs\_GenericOffset} template is used. All other templates, relying on unpredictable \textit{jitter} offsets, will be simply ignored by \textsc{fcmaker}\footnote{Users are urged to keep this point in mind, in the specific (and unlikely) event that a given OB contains first a \textit{jitter-based} template \textbf{without} a return to origin, followed by a \texttt{HAWKI\_img\_obs\_GenericOffset} template. In that case, the finding charts generated by \textsc{fcmaker} will not be correct.}.  The left-hand-side plot is always centered on the acquisition field, with a radius of 310$^{\prime\prime}$. For observations using the fast photometry mode of HAWKI, the finding charts display the detector windowed area. By default, the background image is taken from 2MASS \citep{Skrutskie2006}. The minimum valid radius for telescope guide stars is set to 240$^{\prime\prime}$ from any offset position. 

\paragraph{XSHOOTER}

The finding charts (see Fig.~\ref{fig:XSHOOTER}) show the field-of-view of the acquisition camera in bold purple. The slit, integral-field unit, or acquisition camera field-of-views are then drawn for each offset position (either O or S), according to the selected observing templates. The minimum valid radius for telescope guide stars is set to 120$^{\prime\prime}$ from any offset position. In the case of blind offsets, the target location marked in the X-shooter finding charts is that of the final science target, and not that of the \textit{target} (as defined in the OB). In slit mode, the position angle of the observations is clearly indicated in the top right corner. By default, the background image is taken from DSS2 (red). 

\subsection{Moving targets}
\textsc{fcmaker} is compatible with OBs looking at moving targets, defined either with an ephemeris file attached to the OB, or with proper motions specified alongside the target coordinates. In both those cases, the user can specify the exact date (and time) of the observations to draw accurate finding charts, even for very fast moving objects. The VLT computes the coordinates of moving targets by assuming that they move along a Great Circle on the sky, a suitable approach if the distance to the target is unknown. In \textsc{fcmaker}, the coordinates of moving targets at a given date (and time) is propagated --from the closest ephemeris point in time, if applicable-- using the \texttt{astropy.coordinates.SkyCoord.appl\_space\_motion()} routine from the \textsc{astropy} package \citep{AstropyCollaboration2013}, which assumes instead that the target is moving in a straight line (in space). From the perspective of the finding charts, the error associated with this different treatment of moving targets between \textsc{fcmaker} and the VLT will remain negligible for most cases. Whenever suitable, users are encouraged to provide target coordinates at recent epochs, and/or suitably sampled ephemeris files.

Evidently, the ability to specify the exact observing date (and time) of the finding charts associated with moving targets will be most useful for Visitor Mode and ``delegated'' Visitor Mode observations, or Service Mode programs with tight time constraints. If an ephemeris file is attached to the OB, \textsc{fcmaker} will show the locations of the target within $\pm2$\,hr from the set observing date (see Fig.~\ref{fig:XSHOOTER}). Each ephemeris entry is shown individually, so that no error associated with the extrapolation of proper motions is made on these points. A warning tag, using the symbol ``$\leadsto$'', is automatically added to all finding charts with moving targets (see Fig.~\ref{fig:XSHOOTER}).

\subsection{Determination of parallactic angles}

The ability to compute parallactic angles was built inside \textsc{fcmaker} for instruments that require it, using the \textsc{astroplan} package \citep{Morris2018}. The coordinates and elevation of the telescopes on Cerro Paranal are taken from the ESO website\footnote{\url{https://www.eso.org/sci/facilities/paranal/astroclimate/site.html}, accessed on \today.}. By default, the instrument field-of-view are not drawn on the \textsc{fcmaker} finding charts in case a parallactic angle is requested, since doing so would render the finding chart extremely time-specific. Although this is not recommend for Service Mode observations, users can force the drawing of the field-of-view in all cases with the \texttt{--do-parang} flag. A warning tag, using the symbol ``$\measuredangle$'', is automatically added to all finding charts requesting parallactic angles (see Fig.~\ref{fig:XSHOOTER}).

\section{The future of fcmaker}\label{sec:plans}
Once an instrument at the VLT has been commissioned, the structure of the associated OBs -- including the naming conventions for the observing modes, and the name of the different OB parameters -- are extremely unlikely to evolve. Any modifications to OB conventions would affect sub-systems throughout the entire VLT data flow, and necessarily require a formal commissioning at the telescope. OB conventions may only evolve under specific (and rare) circumstances. Form this perspective, the maintenance of the \textsc{fcmaker} code can be expected to be minimal. It will largely consist of tracking the evolution of the \textsc{python} packages at its core on a biannual basis: the current frequency of ESO's call for observing proposals.

At the time of publication of this article, \textsc{fcmaker} fully supports MUSE, HAWK-I, and X-shooter OBs. Support for UVES and ESPRESSO \citep[Echelle SPectrograph for Rocky Exoplanets and Stable Spectroscopic Observations;][]{Pepe2013} is under consideration. Support for all other existing VLT \& VLTI instruments, including NACO \citep[Nasmyth Adaptive Optics System Near-Infrared Imager and Spectrograph;][]{Lenzen2003, Rousset2003}, KMOS \citep{Sharples2013}, FORS2, FLAMES, SPHERE \citep[Spectro-Polarimetric High-contrast Exoplanet REsearch;][]{Beuzit2008}, VISIR \citep[VLT spectrometer and imager for the mid-infrared;][]{Lagage2004}, SINFONI \citep[Spectrograph for INtegral Field Observation in the Near-Infrared;][]{Eisenhauer2003,Bonnet2004}, OmegaCAM, VIRCAM, GRAVITY \citep{GravityCollaboration2017}, MATISSE \citep{Lopez2014}, PIONER \citep[Precision Integrated-Optics Near-infrared Imaging ExpeRiment;][]{LeBouquin2011}, and AMBER \citep[Astronomical Multi-BEam combineR;][]{Petrov2007} is not foreseen at this stage. Support for future VLT instruments, including CRIRES+ \citep[CRyogenic high-resolution InfraRed Echelle Spectrograph;][]{Kaeufl2004,Oliva2014}, ERIS \citep[Enhanced Resolution Imager and Spectrograph;][]{Amico2012a,Kuntschner2014} and MOONS \citep[Multi-Object Optical and near-infrared Spectrograph;][]{Cirasuolo2014}, may be considered when these get offered to the community. 

These general intentions should however not deter motivated users from contributing to \textsc{fcmaker}. All suggestions and Github push requests for \textit{any} instruments accessible from \textit{p2} will certainly be welcomed, and examined with attention. In all cases, any major evolution of \textsc{fcmaker} from its current state (for example, the support of a new instrument mode) would systematically seek feedback from NAs and TIOs at the VLT, regarding the content and design choices of the finding charts. Doing so will ensure that the code remains true to its spirit: that of the automated creation of ESO-compliant for OBs on \textit{p2}, fine-tuned to meet the needs of night time operations at the VLT.

\section*{Acknowledgements}
{\smaller
I thank ESO Fellows Romain Thomas and Cyrielle Opitom for their help assembling test OBs for X-shooter and moving targets, and Alexandre Santerne for his excellent suggestion to create mock sky images from the Gaia catalogue. I am very grateful to the NAs and TIOs at the VLT for their feedback and suggestions on the design of the \textsc{fcmaker} finding charts. I also thank the two anonymous reviewers of this article for their constructive feedback.\\

This research has made use of \textsc{fcmaker} \citep{Vogt2018a}, a \textsc{python} module to create ESO-compliant finding charts. \textsc{fcmaker} relies on \textsc{matplotlib} \citep{Hunter2007}, \textsc{astropy}, a community-developed core \textsc{python} package for Astronomy \citep{AstropyCollaboration2013}, \textsc{astroquery}, a package hosted at \url{https://astroquery.readthedocs.io} which provides a set of tools for querying astronomical web forms and databases \citep{Ginsburg2017}, \textsc{astroplan} \citep{Morris2018}, \textsc{aplpy}, an open-source plotting package for \textsc{python} \citep{Robitaille2012}, and \textsc{montage}, funded by the National Science Foundation under Grant Number ACI-1440620 and previously funded by the National Aeronautics and Space Administration's Earth Science Technology Office, Computation Technologies Project, under Cooperative Agreement Number NCC5-626 between NASA and the California Institute of Technology. \textsc{fcmaker} uses the VizieR catalogue access tool, CDS, Strasbourg, France. The original description of the VizieR service was published in \citep{Ochsenbein2000}. \textsc{fcmaker} makes use of data from the European Space Agency (ESA) mission Gaia (\url{https://www.cosmos.esa.int/gaia}), processed by the Gaia Data Processing and Analysis Consortium (DPAC, \url{https://www.cosmos.esa.int/web/gaia/dpac/consortium}). Funding for the DPAC has been provided by national institutions, in particular the institutions participating in the Gaia Multilateral Agreement. In particular, \textsc{fcmaker} uses data from the Gaia \citep{GaiaCollaboration2016a} Data Release 2 \citep{GaiaCollaboration2018}.  \textsc{fcmaker} also uses data from the Second Digitized Sky Survey (DSS2). The ``Second Epoch Survey'' of the southern sky was produced by the Anglo-Australian Observatory (AAO) using the UK Schmidt Telescope. Plates from this survey have been digitized and compressed by the STScI. The digitized images are copyright (c) 1993-1995 by the Anglo-Australian Telescope Board. The ``Equatorial Red Atlas'' of the southern sky was produced using the UK Schmidt Telescope. Plates from this survey have been digitized and compressedby the STScI. The digitized images are copyright (c) 1992-1995, jointly bythe UK SERC/PPARC (Particle Physics and Astronomy Research Council, formerly Science and Engineering Research Council) and the
Anglo-Australian Telescope Board. The compressed files of the ``Palomar Observatory - Space Telescope Science Institute Digital Sky Survey'' of the northern sky, based on scans of the Second Palomar Sky Survey, are copyright (c) 1993-1995 by the California Institute of Technology. All DSS2 material not subject to one of the above copyright provisions is copyright(c) 1995 by the Association of Universities for Research in Astronomy, Inc., produced under Contract No. NAS 5-26555 with the National Aeronautics and Space Administration. }



\bibliographystyle{aa}
\bibliography{bibliography_fixed.bib}





\end{document}